\documentclass[12pt]{article}
\usepackage[tmargin=1in,bmargin=1in,lmargin=1.25in,rmargin=1.25in]{geometry}

\usepackage{setspace}
\usepackage{csquotes}
\usepackage{kpfonts}
\usepackage{tikz-cd}
\usepackage{amsmath}
\usepackage{amsthm}

\usepackage{verbatim}
\usepackage{amsfonts}
\usepackage{stmaryrd}
\usepackage{dsfont}
\usepackage{graphicx}
\usepackage[title]{appendix}
\usepackage[font=small,labelfont=bf]{caption}
\usepackage{lipsum}
\usepackage{mwe}
\graphicspath{ {images/} }

\newtheorem{definition}{Definition}
\newtheorem{example}{Example}
\newtheorem{theorem}{Theorem}

\title{Univalence and Ontic Structuralism}
\date{}
\author{Lu Chen\thanks{chen.l@usc.edu}}
\begin{document}
	\maketitle
	
	\section*{Abstract} The persistent challenge of formulating ontic structuralism in a rigorous manner, which prioritizes structures over the entities they contain, calls for a transformation of traditional logical frameworks. I argue that Univalent Foundations (UF), which feature the axiom that all isomorphic structures are identical, offer such a foundation and are more attractive than other proposed structuralist frameworks. Furthermore, I delve into the significance in the case of the hole argument and, very briefly, the nature of symmetries. 
	
\vspace*{4mm}

\tableofcontents

\section{Introduction}

Homotopy Type Theory (HoTT) with the univalence axiom, together known as `Univalent Foundations' (UF), is a novel contender in the ongoing search for the best logico-mathematical foundations for philosophy and science, which is developed by the Univalent Foundations Program (UFP [2013]).\footnote{While `UF' is the abbreviation for a plural noun, the plurality will not be relevant for our purposes (different UF may choose slightly different axioms or rules that do not matter for our purposes). So I will henceforth treat it as a singular noun. Also note that there are different ways in which the term `HoTT' is used, e.g., some people use the term interchangeably with `UF'. } The exploration of its relevance to philosophy is currently limited to mathematical structuralism and its application to the hole argument (see Awodey [2014], Tsementzis [2017], Dougherty [2019], Ladyman and Presnell [2020]). I believe that it is worth delving deeper into the significance of UF, including in the case of the hole argument. But it is particularly important to examine UF's relationship with ontic structuralism, a metaphysical doctrine that roughly says that objects do not have identities over and beyond the structure they occupy. This would help clarify what UF can uniquely contribute to many topics in philosophy of physics, including the hole argument and symmetries. 

But first, what is UF and its univalence axiom?  UF is a radical departure from first-order predicate logic and standard set theory ZF(C) that largely define the landscape of metaphysics. It is a type theory where every entity has a type. Unlike simple type theory, types are part of the object language.\footnote{In simple type theory, types are part of the metalanguage describing syntactical functions of terms in the formal language. In contrast, types are part of the object language in HoTT and presumably refer to a sui generis kind of entities in the world, which can be understood as a sort of structures (but note that the formal notion of `structure' in HoTT/UF is seperately defined; see Appendix B).  For example, I may talk about a type of apples.  Types also have types, which are called `universes'.
	
	According to the standard terminology of HoTT, we say `every term has a type'. This can potentially cause confusion for philosophers, since `term' seems to refer to part of the formal language rather than objects in the world. To avoid the confusion, I simply say every entity has a type. The HoTT theorists may have kept their terminology because of their main focus on formal symbols in mathematics and computer programming. } They are like sets in some respects such as consisting of elements (or otherwise being empty). But they are unlike sets in other respects. For one thing, they can have additional structures in the form of `higher-order' elements (technically, they are conceptualized as homotopy spaces or $\infty$-groupoids, hence the name `homotopy type theory'; see Section 2). The univalence axiom further says that  equivalent types are identical. This can be shown to entail that isomorphic structures are identical (here, the notions of `structure' and `isomorphism' are formally defined in HoTT in a way that captures their more familiar meanings).

The relation between the univalence axiom with mathematical structuralism is clear, since the latter is often defined by that isomorphic mathematical structures are identical. Tsementzis ([2017]) further argues that UF is the only feasible structuralist foundation for mathematics among the existent ones. A natural further step is to argue that UF is also a structuralist foundation for science, where structuralism here concerns physical entities. Thus understood, UF is a foundational implementation of ontic structuralism. This is desirable because ontic structuralism has faced criticisms that it is either incoherent or indistinguishable from its traditional rivals (see McKenzie [2017]). The standard framework's limitations in formulating ontic structuralism as a coherent and distinct doctrine necessitate a transformation in the very foundation. As Sider ([2020]) puts it:

\begin{quote}	
	What basic notions is the structural realist proposing? What are the proposed rules governing those notions? And how can those notions then be used in a foundational account of scientific theories?... You need to properly specify a replacement framework, some replacement inventory of basic notions, rules governing those notions, and methods for using those notions in foundational contexts. (p.64)
\end{quote} 
The core task of this paper is to point out that UF is an answer to all these questions, namely that it is a rigorous logical framework that implements the idea of ontic structuralism and promises to found science. In Section 3, I will compare this approach with two other noteworthy attempts, which are respectively  \textit{generalism} proposed by Dasgupta ([2009]) and \textit{algebraic structuralism} by Dewar ([2019a]). I will argue that they face similar problems. The problems concern their inadequacy to dispense with distinct isomorphic structures and higher-order individuals.  UF does not face these problems.\footnote{\label{elim} As we will see later, both Dasgupta and Dewar try to implement eliminative ontic structuralism, where individuals do not exist at the fundamental level, while UF is non-eliminative in that the objects exist but do not have primitive identities. In this respect, they are quite different approaches from UF. Nevertheless, they are comparable as proposed answers to Sider's questions. See Section 3 for more discussion.} Note that I advance UF specifically as an answer to Sider's questions above, not as a solution to all problems of ontic structuralism. In particular, I am not concerned with the question of what structures we should commit ourselves to given our best scientific theories.\footnote{It is worth noting that ontic structuralism originated as an alternative to (merely) epistemic structural realism, which is a response to the pessimistic meta-induction (see Ladyman [1998]). According to epistemic structural realism, we should only believe in structural features of the theoretical entities posited in our best scientific theories.  There, the central task of structuralism is to discern the structural aspect of a scientific theory that is future-proof and explains the empirical predictions. This is not the focus of this paper. Nevertheless, the proposal here does have some implication for the epistemic task, since we do not need to choose between isomorphic structures.} 

In Section 4,  I will consider and reply to a few objections. One objection is about how we should interpret the notion of identity in the univalence axiom. The other two objections are against ontic structuralism in general: the collapse problem of mathematical and physical structures, and undesirable holism mentioned by Sider ([2020]). The upshot is that UF is at least as resourceful as other structuralist approaches in response to these problems. 


Once UF is viewed as a viable structuralist foundation, its relevance to significant issues in philosophy of physics including the hole argument and symmetries in general  becomes clearer (Section 5). The hole argument, as commonly formulated, relies on the premise that the model $\langle M, g\rangle$ is distinct from $\langle M, g'\rangle$, where $M$ is a manifold, $g$ is a metric field on $M$, and $g'$ is $g$ carried by a non-identity diffeomorphism from $M$ to itself. Ladyman and Presnell ([2020]) have noted that the univalence axiom implies that the diffeomorphically related models are identical, which undermines the hole argument. However, one may ask how this differs from other `mathematical solutions' that also reject this premise.  For example,  Weatherall ([2018]) argues that isomorphic models have the same representational capacity. As he puts it, the hole argument arises from `a misleading use of the mathematical formalism of general relativity' (Weatherall [2018], p.2). However, various authors have pointed out that this does not solve the hole argument, because the latter is about metaphysical possibilities rather than mathematical models (Teitel [2019], Pooley and Read [2021]). To address this, we should generalize mathematical structuralism in the mathematical solution to ontic structuralism.  UF is precisely a structuralist framework for science that blocks the hole argument (which is different from the semantic stipulation imposed by fiat from outside of a formal system in Weatherall's approach).

Additionally, the univalence axiom can be employed to address symmetry-related models more broadly. Various authors have argued that such models should be reformulated as isomorphic ones (see Dewar [2019b]; see also Weatherall [2017] and Wallace [2019]).\footnote{I also attempt to argue this elsewhere in a more nuanced and systematic way.} Under the univalence axiom, these isomorphic models are identified, which enforces that symmetry-related models represent the same physical situation.  This sheds further light on symmetries as result of representational redundacies, albeit very useful ones.



However, given how exotic UF is in comparison with the standard framework, questions naturally arise about whether the benefits of UF are worth the tradeoff with its unfamiliarity.
\footnote{HoTT/UF is gaining independent influence at the forefront of mathematics, particularly in areas like category theory, as an alternative framework to standard set theory and logic. As a foundational language that significantly departs from the standard approach, it is relevant to philosophy, perhaps even in a way similar to how predicate logic and set theory transformed philosophy in the early twentieth century (see Tsementzis and Halvorson [2018], Corfield [2020]). Its significance is not nearly exhausted by the issues discussed this paper and is still very much under exploration and technical development (see for example Schreiber and Shulman [2014]). For instance, synthetic differential geometry is one that can be formulated in HoTT but not in standard logic (see Moerdijk and Reyes [1991]). (Although synthetic differential geometry is standardly formulated in intuitionistic type theory and does not require homotopy types, the point is that the first-order predicate logic is too restrictive for various purposes.)
	
	Nevertheless I should note a limitation at the present stage for discussing UF or HoTT philosophically: it has not been agreed upon \textit{how} UF or HoTT should serve as a foundation for science. Although it is known that we can recover a version of set theory from UF and build science on its basis, such a foundation may not be sufficiently different from the standard one from the perspective of some HoTT advocates (see Section 5). Still, the philosophical significance of the univalence axiom remains unchanged. }  Nevertheless, the exoticness is rather proportional to its significance as an answer to the longstanding tension between our standard logico-mathematical foundation and increasingly structuralist approach to mathematics, science and philosophy.  Given that UF is the only serious structuralist foundation so far, it is not an overtreatment for a small ailment.

\section{An Intuitive Guide to Univalence}

In this section, I will explain the univalence axiom featured by Univalent Foundations (UF) in preparation for the upcoming discussions. My purpose is to help readers understand the univalence axiom (and the involved notions) as efficiently as possible. My strategy is to primarily focus on the semantics based on homotopy theory as a heuristic guide, rather than the syntax. (The syntax  can be rather technical in comparison. For other expositions on HoTT and UF, see for example Shulman [2017b], Ladyman and Presnell [2018], and the main reference book UFP [2013].)

As its name indicates, HoTT not only is a type theory where every entity has a type but also features \textit{homotopy types}. To understand this, let's start with the familiar notion of a topological space, which is invariant under certain continuous deformation such as stretching, shrinking, and bending, but not gluing and tearing. For example, an ordinary cup is topologically equivalent (namely, `homeomorphic') to a simple wedding ring. Every topological space is associated with an algebraic structure called  `$\infty$-groupoid', which contains elements of different levels: its 0-elements are points of the space, its 1-elements are paths between the points, the 2-elements are paths between paths (called `homotopies'), the next level consists of homotopies between homotopies, and so on ad infinitum, without distinguishing elements connected by paths. If we only keep this information, we obtain a `homotopy space'. Intuitively, two spaces are homotopically equivalent if they can be continuously deformed into each other. For instance, a disk is homotopically equivalent to a point. The notion of a homotopy space may sound the same as that of a topological space. Indeed they are closely related, but they are not the same: a homotopical structure typically has less information than the corresponding topological structure; the deformations that the latter permits are more restricted. A disk is \textit{not} homeomorphic to a point (because there is no bijection between the underlying point sets) though they are homotopically equivalent. As another often-cited example, a Möbius strip is topologically distinct from a cylindrical strip, but homotopically equivalent to the latter (both can be deformed into a circle).

Standardly, a homotopy space or a groupoid is defined set-theoretically: a path between two points in a topological space is a continuous map from interval [0,1] to the space that maps 0 and 1 to the two points respectively, and similarly for homotopies. But this does not have to be the case. Instead, we can treat these notions (i.e., paths, homotopies, etc.) synthetically: they can be basic, unanalysable notions directly governed by axioms and inference rules. (As an analogy, analytic geometry reduces  a line to a set of points satisfying various conditions, while synthetic geometry such as Euclid's original formulation of Euclidean geometry treats the notion of line as primitive governed by his basic postulates.) HoTT (or UF) is exactly an axiomatic deductive system where homotopy spaces or groupoids and their high-order elements are basic objects, irreducible to their 0-elements.\footnote{Strictly speaking, it comprises of many axiomatic deductive system variants. The formal system presented in UFP [2013] is one, and there are many others such as univalent cubical type theory (see, for example, Bezem et al [2018]). The paper focuses on the philosophical application and therefore glosses over these technical variants, much like how a big chunk of philosophy of general relativity glosses over different formulations of GR. } The trick to achieve this involves the identity types in HoTT. Suppose $a,b$ are elements of the same type $A$. Then the expression `$a=_Ab$' itself refers to a type, which can have more than one elements, each of which can be thought as a way of identification. These elements can further form identity types, which in turn have elements that can be identified, and so on ad infinitum. This precisely corresponds to the structure of a homotopy space or a $\infty$-groupoid, where `path-connectedness' is reconceptualized as `identification', paths become ways of identification, and higher homotopies become higher-order identifications. In this sense, every type is modelled by a homotopy space---hence the name `homotopy type'.

In other words, homotopy spaces constitute an interpretation for HoTT (or UF).\footnote{More technically, $(\infty,1)$-categories and locally Cartesian closed $\infty$-categories---both are built on the category of $\infty$-groupoids---are respectively the categorical models for UF and HoTT.} Are these interpretations literal or heuristic? That is, are statements in UF really statements about homotopy spaces so that UF is in fact a peculiarly axiomatized mathematical theory? Or, is the homotopy-theoretic interpretation just a tool to prove the consistency of UF and connect UF with classical results in homotopy theory? While I don't have a decisive answer, I can at least say that we are \emph{not} obligated to think UF as just a heuristic reasoning tool for classical homotopy theory. It is perfectly intelligible as an autonomous foundation without appealing to the homotopical interpretations (see Ladyman and Presnell [2018]; also see Tsementzis and Halvorson [2018], Bentzen [2020] for different perspectives). Moreover, the identity types and the univalence axiom are not exclusively motivated  by the homotopy-theoretic interpretation. It is worth mentioning that historically, HoTT did not originate as an axiomatization of $\infty$-groupoids---on the contrary, groupoids are used to informally conceptualize identity types in type theory with more than one elements, which are independently conceived. 

HoTT (or UF) can serve as a logical foundation for science. An important underlying observation is that a version of set theory can be modeled in HoTT  (see Appendix A for more detail). A set is a type that contains no higher homotopical information, called a `0-truncated' type, which has all information beyond its immediate elements discarded. In HoTT,  logical axioms and rules are also subsumed under the rules for type constructions rather than being imposed externally, unlike in set theory. Mere propositions are considered `-1-truncated' type, meaning that it is a set with at most one element (when a proposition is true, it has exactly one element, which is sometimes called its `(truth) certificate', `proof' or `witness', and none if it is false). Standard propositional deduction rules can be derived from type construction rules when the types in question are propositions. In this sense, UF is a unified foundation aiming to replace both first-order predicate logic and set theory. 

Finally, let's turn to the univalence axiom featured by UF. Formally, the axiom says: 

\begin{quote}
	\textsc{Univalence.} For any two types $A,B: \mathcal{U}$, $(A=_\mathcal{U} B)\simeq (A\simeq B)$.\footnote{$\mathcal{U}$ is called a universe, which is a type of types. In the syntax of UF, the expression `$A:\mathcal{U}$' amounts to declaring that $A$ is a type. For curious readers, the quasi-formal definition of $A\simeq B$ is that there is an equivalence $f:A\to B$, where $IsEquiv(f):\equiv (\Sigma_{g:B\to A}f\circ g \sim Id_B) \times (\Sigma_{h:B\to A}h\circ f\sim Id_A)$. Here, `$\sim$' is defined as such: for any appropriate $f,g$, $f\sim g:\equiv \Pi_{x:A} f(x)=g(x)$  (see UFP 2.4 and 2.10). But for our purposes, it suffices to understand that $\simeq$ is a notion defined within HoTT that captures homotopical equivalence.}
\end{quote}
In ordinary English, this says that the identity between $A$ and $B$ is equivalent to the equivalence between them, or in short, equivalent types are identical. This may sound nonsensical, but we can make sense of it using the homotopy-theoretic interpretation. The notion of equivalence between types is precisely that of homotopical equivalence between spaces (except that, of course, it is defined in the language of HoTT rather than classically), and it is a natural idea to identify equivalent spaces.\footnote{On the technical level, it is more delicate. An identity type $A=_\mathcal{U}B$ is inhabited by the paths between $A$ and $B$ as two points in space $\mathcal{U}$ in the homotopical interpretation, while an equivalence type $A\simeq B$ is inhabited by all the maps between $A$ and $B$ as spaces that preserve all their homotopical information. \textsc{Univalence} amounts to saying that any such map can be promoted to a path. Note that the other direction is trivial: according to the principle of `path induction', any property can be transported by a path, so that if $A=_\mathcal{U}B$, and given $A\simeq A$, we automatically have $A\simeq B$. Also note that \textsc{Univalence} is \emph{not} directly justified by the interpretation in terms of standard homotopy theory since the latter does not have the space of spaces analogous to  $\mathcal{U}$.}

		\section{Foundation of Structuralism}

	\textit{Ontic structuralism} can be informally characterized as the view that structure is more fundamental than objects. But it is unclear how the view can be made more precise. Various attempts at it have been criticized as `incoherent' or `indistinguishable from its supposed rivals' (McKenzie [2017]). The incoherence charge is largely against \textit{eliminativism}, an approach to ontic structuralism which says that there are no individuals or objects but only structures at the fundamental level (see alse Footnote \ref{elim}). In contrast, the UF approach that I will propose is non-eliminative.  I propose that the following principle holds, which is understood in the framework of UF:
	
	\begin{quote}
		\textsc{The Identity Thesis (Identity)}. Fundamentally, all isomorphic structures are identical.
	\end{quote}
	This approach does not eliminate objects at the fundamental level: a structure consists of objects and their structural relations (see Appendix B).\footnote{What are objects in this framework? The answer is, roughly, anything. Note that a structure can be built on \textit{any} type (see, for example, Definition 10 in Appendix B).  Unlike most higher-order logics considered in the philosophical literature, there is no basic syntactic type---often denoted by `$e$'---of terms that refer to entities which we ordinarily think of as individuals, e.g., particles, tables and chairs. Here, we can use  any type, which may consist of `higher-order' entities like properties, as the basis of a `higher-order' structure. Indeed, I will argue for the advantage of this feature over the rival proposals that only focus on (eliminating) first-order individuals. }  Rather, objects are intrinsically characterless---they are exhaustively characterized by the structure they are in. Merely permuting which objects instantiate what relations in a structure would not lead to any real change, but merely notational.
	So we can consider this approach as a version of \textit{priority-based structuralism}, according to which both structure and objects exist but the former is more fundamental than the latter.\footnote{Note that this claim is compatible with the denial of eliminativism. Structures are more fundamental than objects even though objects also exist at the fundamental level. As McKenzie [2007] puts it, `structuralists are discerning metaphysical structure \textit{within} that most fundamental level, arranging the categories of entities that feature there into relations of ontological priority.'(p.5; italic original) } Indeed, we can think of types as structures of a minimal kind, and they are more fundamental than their elements.\footnote{In Appendix A, I explain the UF-theoretic notion of sets and how---according to it---sets are more primary than their elements. It is thus unsurprising that a structure, which is built on types, is more fundamental than the elements. See also the discussion in Section 4.2 on variables, which in UF behave like constants for entities without haecceities. }

I will advance UF-implemented \textsc{Identity} as a response to Sider's ([2020]) complaint that ontic structuralism lacks a rigorous logical framework. In particular, I will compare it with  \textit{generalism} by Dasgupta ([2009]) and \textit{algebraic structuralism} by Dewar ([2019a]), both of which are explicitly proposed as formal frameworks for ontic structuralism, and both are eliminativist.  I argue that UF avoids the problems with these frameworks.

 \subsection{Generalism} 
Generalism is based on predicate functor logic developed by Quine ([1976]) and others, the syntactical categories of which do not include variables or constants for individuals. Instead, predicates are considered atomic terms. For example, $L^2$ replaces the standard formula $Lxy$ (`x loves y'), where the superscript indicates the number of arguments of its standard counterpart. Furthermore, we have a handful of predicate functors that operate on predicates (including standard logical connectives $\neg$ and $\wedge$). For example, if we want to say $\exists x\exists yLxy$ in standard logic, we simply say in predicate functor logic `$cc L$', where $c$ is a functor that behaves as if it binds the first free variable in the formula in its scope. As another example, the standard formula $\forall x(Cx\to\exists yLxy)$ (`every cat loves something') can be expressed as $\neg c(C^1\wedge \neg c\neg \sigma L^2)$, where the operator $\sigma$ permutes the arguments of a predicate (for otherwise the expression would mean `every cat is loved by something'). Predicates and functors are the only syntactical categories other than logical ones.

Importantly, we can show that algebraic generalism is expressively equivalent to  first-order predicate logic without constants, which is called \textit{quantifier generalism} (see Appendix in Dasgupta [2009]). This is not surprising, since predicate functor logic is designed to be an algebraization of first-order predicate logic. The translation scheme between the two systems is laid out in Quine ([1976]) and other authors. Axiomatic set theory ZFC, the standard foundation for mathematics, is expressible in first-order predicate logic, and is therefore expressible in predicate functor logic or generalism. In this sense, generalism can serve as a foundational framework for science (albeit not convenient to use---the interested readers can try translating the axiom of extensionality).

 Consider a model $\langle D, v\rangle$, which is a structure, where $v$ is a set of relations defined on elements of $D$. We say that  $\langle D, v\rangle$ and $\langle D', v'\rangle$ are isomorphic if there is a bijective map between $D$ and $D'$ that preserves all the relations.\footnote{\label{iso}There is a potential ambiguity in the phrase `preserving all the relations' depending on one's stance on what relations are. On the purely extensional reading of relations, a model where the only two things `love' each other would be isomorphic to one where the only two things `are next to' each other. On the intensional reading, clearly they are not isomorphic, since the relations are different and thus not preserved. In this paper, I generally refer to the latter sense of isomorphism. Which approach is used in standard mathematical practice? The answer is that both can be useful. We normally require an isomorphism to be between objects of the same type with the same signature (e.g., between two groups, or vector spaces, or graphs, etc).  But we also talk about isomorphisms between structures of different types or with different signatures (e.g., we say that $\langle \mathbb{Z}/7-\{0\}, \cdot\rangle$ and $\langle \mathbb{Z}/6, +\rangle $ are isomorphic).} Under generalism, we can still define distinct isomorphic models without constants (even as part of normal practice).  For instance, let $D$ consist of natural numbers 0 and 1 and $v$ consists of a single relation holds of 0 and itself alone. We can easily define a distinct isomorphic structure $\langle D, v'\rangle$ where $v'$ consists of the same relation holding of 1 and itself alone. Note that we do not need constants to express natural numbers, since they are definable in standard set theory as different sets (such as through von Neumann reduction). Such examples are abundant. In general,  the apparatus of set theory allows us to represent the elements of a structure in many different ways. Algebraic generalism has the expressible power of set theory in standard logic, thus it is capable of distinguishing isomorphic structures that generalists aspire to identify.

\subsection{Algebraic structuralism} 

Dewar ([2019a]) complains that generalism does not really dispense with individuals because the semantics for the framework is the same as that for the first-order predicate logic.\footnote{Dasgupta defends this approach by arguing that the semantics is dispensable and the syntax alone is sufficient for understanding a theory. Dewar is not convinced.} That is, we use models $\mathcal{M}=\langle D, v\rangle$ to interpret sentences formulated in generalism, where $D$ is the domain of individuals. 


As an alternative, Dewar  proposes \textit{algebraic structuralism}.  Instead of eliminating syntactical categories of variables and constants for individuals, Dewar seeks to eliminate the individuals in the semantics of theories (I'll come back to comment on this transition from syntax to semantics).  Again, consider a standard first-order model $\langle D, v\rangle$.  In the set-theoretic framework, all the relations in $v$ are definable as set-theoretic constructions based on $D$. Now, $D$ is a domain of individuals, which we want to get rid of. Can we reconceptualize the relations in $v$ without resorting to elements of $D$? Yes. For simplicity, suppose $v$ has only monadic properties, which are construed as subsets of $D$.  Now, we can reconceptualize those properties as elements of a Boolean algebra defined by algebraic operators including disjunction (or `addition'), conjunction (or `multiplication'), and negation. These operators are governed by axioms of Boolean algebras rather than defined through set-theoretic constructions. From such a Boolean algebra, we can recover the set-theoretic definitions of monadic properties, so we can ensure that this reconceptualization does not lose information. This idea can generalize to arbitrary first-order model  $\langle D, v\rangle$ where $v$ has more than monadic properties. The resulting algebras are called `cylindrical algebras'. Shortly put, Dewar's algebraic structuralism amounts to an algebraization of standard first-order models. 

In philosophy of physics, this idea is implemented more directly through algebraizing relevant models, such as spacetime models. Instead of taking manifolds as fundamental and defining physical fields on them (which can be considered as distributions of qualitative features over spacetime points), we can take fields as primitive entities without an underlying manifold---call this \textit{algebraicism} (see Geroch [1972], Rosenstock et al [2015], Chen and Fritz [2021]). This works because the fields can encode all necessary information about manifolds for doing physics up to general relativity and arguably also quantum field theory.

However, adopting algebraic structuralism or algebraicism that dispenses with spacetime points does not solve the problem of individuals in principle, nor do they satisfy \textsc{Identity}. Under algebraic structuralism, models are now in the form of $\langle R, O\rangle$, where $R$ is a set of properties and relations and $O$ is a set of operators on them. But this is formally the same as $\langle D, v\rangle$, and the difference mainly concerns whether $D$ contains ordinary individuals or ordinary properties and relations (or rather tropes since we are really talking about instances of properties and relations). The problem is most salient in algebraicism where physical fields are the new individuals. It is long observed in the literature that we can just as easily come up with isomorphic models consisting of physical fields (see Rynasiewicz [1992]).\footnote{It is often argued that algebraicism is equivalent to the standard manifold-theoretic approach.  I argue elsewhere that they are not equivalent in philosophically important ways. But this discussion is tangential to this paper.} As another example, suppose being negatively charged and positively charged are completely symmetric in their nomic roles so that exchanging them has no empirical consequences. Since they are expressed as distinct members of the domain in $\langle R, O\rangle$, the algebraic formalism still entails that exchanging them results in a distinct physical situation.  Thus, while there are no ordinary individuals in the algebraic models, the problem of individuals still arises. 

Of course, we can try to remedy this problem by algebraicizing models at the second order. But even if this attempt is technically feasible, it does not guarantee that the problem does not arise for the higher-order entities and thus cannot solve the problem once and for all.\footnote{Does this mean that thoroughgoing structuralism is in principle impossible because this `structuralizing' process should go arbitrarily high-order?  If so, this would be bad for the `high-order' argument against algebraic structuralism as long as one holds that structuralism is possible in principle. But I think the answer is no---the process can legitimately end somewhere for structuralists. Consider the simple theory that contains only one statement $P(a)$. Although $P$ and $a$ are primitive notions, the theory does not violate the spirit of structuralism, since there is nothing to permutate that would result in empirically equivalent theories (exchanging $P$ and $a$ is not allowed since the result would not be well-formed). Or consider a theory with only the following statement: there is an x and there is a y such that x is next to y and neither x nor y is next to itself. The structuralists won't object to this theory even though `next to' is a primitive notion not reduced to any higher-order structure. Thus it is in principle possible for structuralists to posit some primitive nonstructural notions.} Indeed, this problem is recognized and discussed by Dewar himself, who concedes that there are potential higher-order problems, but adds that eliminating first-order individuals is still a progress for ontic structuralists. 

Note that algebraic structuralism and generalism are in the same boat regarding my criticisms. The criticism of generalism also applies here, namely that as long as the models are formulated in standard set-theoretic foundation, we cannot rule out isomorphic but distinct models. The problem of higher-order individuals also applies to generalism, since it does not allow invariant permutations of predicates. In contrast, UF does not face these problems since UF-implemented \textsc{Identity} applies to structures of any order. 

\paragraph{Comments on Syntax vs Semantics} When criticising Dasgupta's approach, Dewar writes:

\begin{quote}
What is needed is a semantics for G which explains how the world could make sentences of G true, even if the world is [...] not fundamentally constituted by individuals standing in properties and relations[.] [M]etaphysical proposals should address themselves to semantics not to syntax. (Dewar [2019a], p.1845)
\end{quote}
I agree with Dewar that we should be able to say how the world is such that the sentences in our best theory are true. This, in Dewar's context, amounts to a model theory without individuals. Since UF is advanced as our fundamental framework, it is natural to use it to not only regiment our physical theories but also explain how the world makes them true. It is adequate for this purpose (recall that a version of set theory that satisfies \textsc{Univalence} can be recovered from HoTT, which is powerful enough for such a model theory; see Appendix A). Indeed, the main application of the UF-implemented ontic structuralism, as we will see in more detail in Section 5, is to clarify the relation between symmetry-related models for our physical theories.  

At the meantime, I want to point out that  there is no clear distinction between a model-theoretic representation and a sentential one in the UF formalism unlike in standard logic, so we should not assume such a dichotomy when we switch to the UF approach. In standard logic or simple type theory, a model is an individual that can be predicated of while a sentence is not. So there is a clear syntactical distinction between representations of a worldly structure and a proposition. In contast, in UF (or HoTT), sets and propositions are both types and belong to the same universe of types.  Indeed, a proposition is a set with at most one element. But even this formal difference is not conceptually important for us.  Indeed, an ordinary existential statement can be directly formalized as a set with more than one element in HoTT, where the elements are its truth witnesses, and only then is truncated to a proposition by identifying all the elements by fiat. 

Let's briefly consider a concrete example.  The theory of general relativity can be written out as the Einstein field equations, which are statements about how the metric field on the spacetime manifold is correlated with the matter distribution---denote them by `EFEs'. In HoTT, since propositions are types, we formalize them in the form of `$a:$ EFEs' (see Section 4 for more details of the syntax).  Now, consider solutions to EFEs, which are set-theoretic models. If we assume that these solutions are fully characterized by the field equations, they can be formalized in HoTT in the form of `$z:$ SolvesEFEs', which gives the models that satisfy EFEs, with `SolvesEFEs' denoting the type of such models (see also Section 5).\footnote{Less elliptically, this can be written out as $z: \Sigma_{M:\textrm{Mfd}}\Sigma_{g:\textrm{Metric}(M)} \textrm{SolvesEFEs}(M,g)$, which can be further unpacked. } 
We can see that there is little difference between these two ways of describing our general-relativistic world (again, the only formal distinction between them is the number of elements, which---as I have contended---plays no important conceptual role).\footnote{ One can imagine this position to be further developed (though beyond the scope of this paper), perhaps in connection with similar approaches in the literature of higher-order logic.  See Bacon ([2023]).}

\section{Objections and Replies}

I will turn to some possible objections to UF-implemented ontic structuralism, some of which do not specifically target at the UF approach but are helpful for clarifying features of UF.

\subsection{Interpretation of Univalence} 

Although UF formally implements \textsc{Identity}, there is a glaring interpretative issue of the identity type. One may object that the identity in UF really means indiscernibility instead, and therefore we haven't said anything new by the univalence axiom---surely, isomorphic structures that preserve all observable structures are indiscernible. This reading is especially consistent with the homotopy-theoretic interpretation. Recall that when we say $a=_A b$ ($a,b$ are elements of type $A$), $a,b$ can be interpreted as two points in a homotopy space $A$ that are connected by a path. These two points are topologically indistinguishable but need not be one and the same point.  Nevertheless, as I have commented, the homotopy-theoretic interpretation is useful to provide intuitions for UF but the latter can be perfectly understood without appealing to the former. Indeed, I shall argue that identity should be understood literally as identity. Assuming that the notion of indiscernibility between structures is captured by the notion of isomorphism, we can say that \textsc{Univalence} successfully implements the principle of the identity of indiscernibles. My reason is simply that identity in UF is defined in the exact same way as in standard logic, barring syntactical differences. In standard logic, identity is defined as follows:

\begin{quote}
	A binary relation $=$ is an identity relation if (1) $(\forall x) x=x$ is an axiom or a theorem; (2) for any well-formed formula $\mathcal{F}$, we have as an axiom or theorem that 
	$$\forall x\forall y(x=y\to (\mathcal{F}[x,x]\to \mathcal{F}[x,y]))$$ where $\mathcal{F}[x,y]$ is obtained from $\mathcal{F}[x,x]$ by substituting zero or more free occurrences of $x$ by $y$.
\end{quote}
The definition in HoTT is completely analogous:
\begin{quote}
	For any type $A$, the type $=_A$ is a binary relation that satisfies: for any $x:A$, (1) $x=_Ax$; (2) for any type $P$, we have $$ind: \prod_{(y:A)} (x=_Ay\to (P[x,x]\to P[x,y])) $$ where $ P[x,y]$ is obtained from $P[x,x]$ by replacing any number of occurrences of $x$ by $y$.

\end{quote}
Here, $`\prod'$ behaves like a universal quantifier. (2) is directly entailed by \textit{path induction}.\footnote{Embedded in this context, path induction can be briefly expressed as $$ind: \prod_{(y:A)} \prod_{(p:x=_Ay)} (P[refl,x,x]\to P[p,x,y]) $$} Barring technical details that are irrelevant to the current discussion, these two definitions are exactly the same. One may think that the univalence axiom alters the meaning of the identity type, but that should not be the case, since the above conditions are \textit{sufficient} for identity in standard logic.\footnote{Some authors have pointed out several features of the identity type that correspond to Leibniz's principles for identity (nLab [2023a]). Apart from the ones mentioned in the main text, they also mention a third feature: an identification identifies itself with self-identification. This, I think, is uniquely HoTT, and not a counterpart of Leibniz's theorem `quae sunt eadem uni tertio, eadem sunt inter se' as opposed to what the authors claim (Gerhard [1890], p.230).} 

Note that \textsc{Identity} does not lead to an identification of merely weakly discernible individuals, which are qualitatively identical. Two things are \textit{weakly discernible} if there is an irreflexive relation that holds between them (see Ladyman and Ross 2007). A typical example is a toy universe where the only two existents are two symmetrically-arranged half-disks---they are weakly discernible because they are `next to' each other. What's more, \textsc{Identity} does not lead to an identification of two identical particles which have no qualitative or structural difference from each other (in particular they don't have distinct spatiotemporal positions, unlike the half-disks). A type that contains two elements can be distinguished from that with one element based on the number of its self-identifications (see Appendix A). To emphasize: the permutation invariance of the elements of a type does not entail the collapse of all elements into one. Endorsing \textsc{Identity} does not lead to the conflation between strict identity and qualitative identity.

It might be worth mentioning---though this has been relatively well discussed---that in UF or HoTT, in addition to the identity, we also have a notion of judgmental equality (denoted by `$\equiv$').  If two terms or types are judgmentally equal, say $A\equiv B$, then they are interchangeable in any circumstances within HoTT.  But if we just have $A=B$, then we cannot replace $A$ with $B$ in an expression like $`a:A'$.\footnote{\label{internal}Note that this does not violate path induction. Judgments like `$a:A$' or `$A\equiv B$' does not have a type and cannot be embedded in a more complicated statement: for instance, it is syntactically ill-formed to say $\neg (A\equiv B)$. So they do not fall under the scope of path induction. In this sense, judgmental equality like $A\equiv B$ is not fully internal to the formal system. However, it is internal in the sense that it is governed by inference rules in HoTT like other expressions.} Indeed, if we could exchange $A$ with $B$ freely, then substituting $A$ with $B$ in the trivial judgment $A\equiv A$ would immediately give us $A\equiv B$, and it would follow that identity entails judgmental equality.\footnote{\label{uip} This is in fact true if we assume \textit{uniqueness of identity proofs} (UIP) in an extensional variant of HoTT. However, UIP is incompatible with the univalence axiom. This is roughly because the rule amounts to a truncation of HoTT into set theory, which in turn is incompatible with the univalence axiom. Assuming UIP, all types are 0-truncated, which means that---to recall---any two elements has at most one way of identification. Such types are set-like and called `h-sets'. The univalence axiom entails that not all types are h-sets.  (See nLab [2023b])
	
Also note that the fact that the identity between elements holds in virtue of the types they inhabit does not mean that their identity is `relative' in the sense that it holds in one way but not in another way. This is because every element has a unique type. The identity (or nonidenity) between two elements is absolute, and holds in virture of the unique type that they inhabit (see also Ladyman and Presnell [2017]).
	
		\label{incomp}	The reason that set theory is incompatible with univalence is straightforward: it would follow from the univalence axiom that all sets with the same cardinality are identical, 	which conflicts with $\{\emptyset\}\neq \{\{\emptyset\}\}$ (Angere [2021]).}	(If such an substitution is possible, the system is called \textit{extensional}. UF, which denies this, is \textit{intensional}.) However, this does not mean that judgmental equality is a better candidate for expressing strict identity in HoTT than the identity type. Indeed, I endorse the standard interpretation of the judgmental equality as indicating the synonimity of two expressions.\footnote{But there are objections to this (see for example Bentzen [2022]).} It is a peculiar but important feature of UF that we can keep a record of distinct expressions within the formal system that refer to the same entity (see Footnote \ref{internal}).

\subsection{Undesirable holism?}

Sider ([2020]) objects to generalism (recall: formalism without constants for individuals) that it leads to holism in the sense that we need to use a world-sentence that describes the world, which cannot be broken down to smaller units (since without constants, we cannot form atomic sentences like $Rab$, and the whole world structure can be relevant to any entity in it). This strikes him as undesirable since it seems alien comparing to our cognition that focuses on localized bits of our world and is in this way economical. But this crticism assumes a distinction between how we use variables and constants that does not apply to formalisms that use sequent calculus including UF. I shall use this as an opportunity to explain how variables in UF behave like constants and yet are not constants in the ordinary sense. For example, an expression that contains unbound variables can be a complete unit of information. A free variable can also refer to the same thing across different formulas. 

Let's look at the relevant details of the syntax of UF to see how this works. A \textit{judgment} is a building block of inferences in HoTT and takes the form $\Gamma \vdash a:A  $ where $\Gamma$ is called a \textit{context}, which contains conditions for $a:A$. An \textit{inference} is of the form:
\begin{quote}
\begin{singlespace}
	$\Gamma \vdash a:A  $
	
	$\rule[0.05in]{1in}{0.5pt}$
	
	$\Gamma' \vdash b:B  $
\end{singlespace}
\end{quote}

\noindent An inference can also have more than one premise.\footnote{This is just like sequent calculus for traditional logic, where every judgment itself is an inference from premises to a conclusion. Thus, an inference or proof in sequent calculus can be considered as a reduction (from bottom to top) of a valid inference to simpler inferences.}  As an example, let's regiment the famous syllogism `Socrates is a man, and all men are mortal; therefore Socrates is mortal' in UF. Let $\Gamma$ be a context including `Socrates: Human, Mortal: Human$\to$ Proposition, z: $\prod_{y:Human}$ Mortal(y)'. Note that `Socrates' is a variable (see the discussion below).\footnote{Here, we simply declare Socrates to be a human without further information. But we can add  any information we associate with Socrates into the context $\Gamma$, such as being Plato's teacher.} 
\begin{quote}
	
	\begin{singlespace}
		
		$  \Gamma \vdash Socrates : \textrm{Human} \qquad\quad \Gamma\vdash z:\prod_{y:Human} \textrm{Mortal}(y) $
		
		$\rule[0.05in]{4in}{0.5pt}$
		
		$\Gamma \vdash z(Socrates): \textrm{Mortal}(Socrates)$
	\end{singlespace}
\end{quote}
\noindent The first premise says that  Socrates is a human in the context $\Gamma$, and the second premise says that all humans are mortal (in $\Gamma$). These premises are very simple valid inferences, since their right-hand sides are part of the left-hand sides. The conclusion is that Socrates is a human (in $\Gamma$), which follows from the premises by $\prod$-elimination, which is like universal instantiation (UFP, A2.4).  

What is the relevance of all this? We observe that a statement (i.e., the right side of a judgment) can have unbound terms and variables as long as they are declared in the context. For example, $Socrates, z $ and even $Human$ are all variables (only variables can be introduced by a context). Also, these variables can appear across different statements within the same proof without changing their references and meanings. Thus, there is no need for a constant `Socrates'. Indeed, it would be very unnatural to introduce `Socrates' as a constant in UF.\footnote{To introduce a constant, we can only do so while defining an inductive type. For instance, when defining the type of natural numbers, we can introduce $0$ as a constant and every other natural number is constructed upon $0$.} Now, one might argue that these variables behave exactly like constants in standard logic, and therefore they \textit{are} constants. To respond, these variables, while like constants in the familiar sense in some respects, are still importantly different from them because `$a:A$' refers to anything that satisfies $A$ and does not presume primitive identity. If we declare $a,b:A$, then exchanging them in any statement preserves its truth value by the inference rule in UF (or HoTT). We cannot say the same about the constants in standard logic: substituting constants by each other is not allowed unless they are explicitly identified. \textsc{Univalence} further reinforces that the different notations we use for isomorphic types and terms have no ontological significance. In this sense, we can say that variables in UF are constants for entities without haecceities.

\subsection{The collapse problem}

Let's start with the notion of `physical entities' as referring to concrete things that we can be in contact with (typically understood as having spatiotemporal locations or being parts of spacetime) and `mathematical entities' as abstract things that exist in some Platonic heaven. Accordingly, mathematical structures are abstract while physical structures are instantiated by concrete things, e.g., the physical structure of two particles next to each other exist if and only if there are two particles next to each other. Now, there is an objection that, if we assume \textsc{Identity},  then the isomorphic mathematical and physical structures will be identified. This is sometimes called `the collapse problem' (see French [2014]). Given the abundance of mathematical structures, for any physical structure, we can find an isomorphic mathematical one.   So it seems that ontic structuralists who adopt \textsc{Identity} must systematically conflate these two kinds. I think this worry is not as formidable as one might think, but a thorough solution to this problem requires a detailed account of \emph{how} we formulate our physical theories and model physical systems within UF, which exceeds the scope of this paper.\footnote{My reply in this section focuses on arguing that UF does not introduce \textit{new difficulties} for structuralists regarding the collapse problem; on the contrary, it has ample resources for implementing various structuralist strategies for answering the challenge. But I will not discuss what strategies are best for answering the challenge, which requires an in-depth discussion that exceeds the scope of the paper and distracts from the main discussion. } But I will very briefly sketch some toy models and strategies just to illustrate why the  UF approach does not need to lead to  a problematic collapse of physical structures into mathematical ones.

 Let's first recall how isomorphism is usually used in standard mathematics: an isomorphism is a structure-preserving mapping between two structures \emph{of the same type} that can be reversed by an inverse mapping (see also Footnote \ref{iso}). For example, we talk about isometries between metric spaces, and homeomorphisms between topological spaces, but not between different kinds of spaces. In UF, this is not only true but highlighted. Like identity claims,  the notion of isomorphism is only defined relative to certain types. Consider $a,b:$ Met, where Met is the type of metric spaces.  If $a,b$ are isometric, then they are identified through \textsc{Univalence}, namely $a=_{Met}b$. Here, isometry is precisely defined relative to the type of metric spaces.\footnote{More detailedly, see Appendix B for the definition of a structure. The identification of isomorphic structures  is entailed by `the structural identity principle' (UFP [2013], \textsection 9.8; see also Appendix B and Section 5).} This is not surprising, for otherwise it would lead to an ill-defined identity claim. 

How do we distinguish between (say) a mathematical manifold and a physical one that are allegedly diffeomorphic, provided that we want to? The answer is to be expected: we can do so by distinguishing between the mathematical type and the physical type, in which case there can be no isomorphism between their elements.  Of course, from the point of view of structuralists, the distinction had better not be primitive but based on structural differences. The merit of UF is that it has ample resources to implement whatever structuralists come up with. For example, we can let the physical type incorporate an observer.\footnote{Formally, this can be as simple as singling out a spacetime point via a 0-ary operator, representing the position of the observer.} Or we can let the physical type incorporate a causal structure (French [2014]). 
We can even make a more systematic distinction by delineating between a physical `universe' (or `kind') and mathematical one (here, `universe' or `kind' refers to higher-order types; see Pierce [2002], \textsection 29). 

Moreover, even if we do not use any of these strategies, we can still incorporate both physical and mathematical structures into the same global model, and straightforwardly distinguish between them by singling out what actually obtains.\footnote{Note that here we are talking about the model of the entire world. If we merely model a proper subsystem, it is frequently the case that the same model can refer to multiple numerically distinct systems (e.g., consider the two half-disks). In this case, it would not be surprising or problematic if a model conflates a mathematical and a physical structure. 
	
	Note that it is a separate worry that the primitive notion of `being instantiated' or `obtain' is not sufficiently structural. This is not a worry against but from structuralists.} 
	
To illustrate, let's consider the approach of algebraicism (see Chen and Fritz [2021]; also mentioned in Section 3), and suppose that the universe is so simple that it contains just one scalar field. Using the apparatus of Einstein algebras, we can model such a universe by an algebra $ C^\infty (M)$ that consists of all smooth functions on a manifold $M$ defined by their smooth operators (see Geroch [1972]). Each element represents a scalar field configuration. Now, regarding the ontological inventory of the world, it is intuitive to think of the actual field configuration as physical and all the other configurations as mathematical. In such a model, we can single out the physical field configuration by stipulating which obtains. The resulting model looks like $\langle C^\infty (M), \{\psi\}\rangle$, where $\psi$ is an algebraic 0-ary operator (i.e., a constant) representing the physical field configuration in $ C^\infty (M)$. 

Of course, it may not even come to that. Perhaps we should simply reject the need to be realists about \emph{both} physical and mathematical entities that are structural identical.
For example, we may `collapse' physical and mathematical structures by endorsing physicalism about mathematical entities (see, for example, Maddy [1990]; for related strategies, see French and Ladyman [2003], Tegmark [2006]). At the meantime, we may endorse nominalism (or anti-realism) about mathematical entities that cannot be plausibly construed in this way. Arguably, both physicalism and nominalism are attractive independently of ontic structuralism. In a sense, the aforementioned algebraicism can be interpreted as adopting this strategy. Since $\langle C^\infty (M), \{\psi\}\rangle$ is a model of our world, there is a sense that the whole model is physical and \textit{every} constituent of the model is also physical. A alleged mathematical configuration can be considered as an actual but uninstantiated state or property of a physical field. In this sense, we effectively rebrand what we originally considered mathematical (namely unactualized field configurations) as physical.

The upshot is that, once we look at the technical details of how we would like to model our physical world in UF together with various structuralists' strategies, we would see that there is no special difficulty regarding the collapse problem at all for the UF advocates. We can implement numerous strategies structuralists have advanced and can advance.

\section{Revisiting the hole argument}

Let's put UF in action. It would be instructive to see how UF applies to the hole argument, since spacetime points are considered exemplary structural entities, and how this approach fares comparing to other solutions to the hole argument. 

First, let's see how UF tackles the hole argument---this is not straightforward in its technical aspect. The hole argument (as it is standardly formulated) relies on the premise that diffeomorphically related models $\langle M,g\rangle$ and $\langle M,g'\rangle$ represent distinct physical possibilities, where $M$ is a smooth manifold, and $g$ and $g'$ are two metrics on $M$ related by a diffeomorphism from $M$ to itself. One might expect that, since  these two models are isomorphic, \textsc{Univalence} directly entails that they are identical, which blocks the hole argument. But this is not the case. First of all, \textsc{Univalence} only applies to types, but $\langle M,g\rangle$ is not a type, but \textit{has} the type of Lorentzian manifolds (it cannot be a homotopy type itself since it includes more structure than its elements and their homotopical information). Also, while I have informally explained the notion of equivalence between types, the notion of equivalence between structures is not even defined.

Fortunately, this can be solved satisfactorily via  `the structure identity principle', which identifies isomorphic structures given \textsc{Univalence}  (UFP [2013], \textsection 9.8; Appendix B). Recall that we can recover a version of set theory from UF, so defining a manifold as a set-theoretic structure is not a problem in UF. We can thereby recover the standard definition of manifold in UF, namely that a manifold is a set of points equipped with topological and differential structures (technically, an `atlas').  The structure identity principle entails that we can identify such set-theoretic structures that are isomorphic (Appendix B). While the full technical detail is quite involved, this result is rather intuitive. Recall that a set is a 0-truncated type (Section 2). \textsc{Univalence} entails that any two equinumerous sets  are identical, with every bijection corresponding to an identification. Given two structures based on a set, it is intuitive that every self-identification of the set that preserves all the structure is an identification between the two structured sets (or rather, a self-identification of the structured set). Thus, \textsc{Univalence} entails that two isomorphic structures are identical. With this in mind, we can indeed easily show that the diffeomorphically related models are identical, as long as we model the metric $g$ as a structure dependent on the type of manifolds $M$ in $\langle M,g\rangle$, which is only natural (see Ladyman and Presnell [2020] for more details).  The same kind of reasoning can show that more complicated diffeomorphically related models with additional matter fields are also identical.\footnote{The downside of this approach is that it is mediated by set theory, and as a result it is more conservative than the aim of UF in serving as a new foundation than many UF advocates intend. 
A more radical approach is to extend UF in a way that manifolds or topological spaces can be directly conceptualized as types. Recall that types already have richer structures than sets in containing higher homotopical information. Now, we can augment UF so that a type not only carries homotopical structure but also additional topological or even smooth structures. Thus, just like the current UF or HoTT is a synthetic theory about homotopy spaces, the extended UF or HoTT is a synthetic theory of geometry and topology. How do we achieve this? Just as the homotopical interpretation of types is justified by the rules and axioms that govern the types in HoTT, the desired interpretation can be achieved by augmenting the rules and axioms that govern the behaviors of topological spaces or smooth manifolds (for the topological version of HoTT called `cohesive HoTT', see Shulman [2017a]; for the geometric version of HoTT called `differential cohesive HoTT', see for example Cherubini [2018]). In the resulting theory, it would directly follow from \textsc{Univalence} that diffeomorphic manifolds are identical.}

Now that I have laid out the univalence solution to the hole argument, we can ask what exactly this solution offers to the existent literature. It falls under the camp called `mathematical solution' which generally argues that there are no distinct diffeomorphically related models in the hole argument (in contrast,  metaphysical solutions are those that argue distinct diffeomorphically related models represent the same physical situation). In particular, Weatherall ([2018]) argues that isomorphic models have the same representational capacity.  Does the univalence approach offer anything new? 

Let's start by reviewing Weatherall's rebuttal of the hole argument, which says that the mathematics of general relativity `does not force one to confront a metaphysical dilemma' contrary to what the hole argument purports to show (Weatherall [2018], p.16). Once again, consider two diffeomorphically related models $\langle M,g\rangle$, $\langle M, g'\rangle$, which are isometric. It is clear that they are isomorphic with respect to an isometry $\phi$. (To be more precise, Weatherall distinguishes between the automorphism $\phi$ and the isomorphism $\phi'$ induced by $\phi$, with the former a map on $M$ and the latter between the two models. But I will gloss over this distinction.) However, when we further infer that the two models are distinct, namely that $g$ and $g'$ assign different metric properties to some spacetime points in $M$, we implicitly appeal to the identity map on $M$.  But the identity map does not induce any isometry: it does not take $g$ to $g'$. So the two models are not isometric or isomorphic \textit{relative to} the identity map. That is, if we stick with the identity map as the relevant comparison, then the two models are neither mathematically equivalent (nor empirically indistinguishable, if we let an observer be located at the same point with respect to the identity map). The models are only equivalent and indistinguishable with respect to isometries such as $\phi$. So when we invoke the hole argument, we no longer consider the identity map as the relevant comparison on $M$. Thus, from the mathematical point of view, it is not true that $g$ and $g'$ assign different properties to the same spacetime points.

As Weatherall adds, if we want to keep to the identity map while comparing the isometric models, we are in fact claiming that there is some additional structure that is not preserved by an isometry, which amounts to saying that a model of general relativity is not a Lorentzian manifold.

There are various authors objecting to responses along this line. For example, Pooley and Read ([2021]) argue that Weatherall's arguments are faulty and leave the hole argument largely untouched.  Contrary to what Weatherall claims, the hole argument does not involve an `illegitimate equivocation' between the identity map and isometry $\phi$---both are invoked for their legitimate purposes.  In particular, when we compare two models $\langle M,g\rangle$ and $\langle M, g'\rangle$ relative to the identity map, we do have two distinct physical situations (different properties are assigned to the same points) that are empirically indistinguishable (since the models are isomorphic simpliciter). To illustrate this point, Pooley and Read give a simple analogy. Alice and Barbara are twins who can look identical if they want. In one situation, Alice wears a red hat and Barbara blue, and they are otherwise indistinguishable. In another situation, they have exchanged their hats and positions so that it looks exactly like the first situation.  In this case, it would be absurd to say that we can only either claim that the two situations are compared relative to the identity map and are empirically distinguishable, or that they are compared relative to an isomorphism (which assigns Alice to Barbara and Barbara to Alice) and are not physically distinct. But this is analogous to Weatherall's claim. 

Putting the objection differently, when we compare $\langle M,g\rangle$ and $\langle M, g'\rangle$  with respect to isometry maps, the self-identity of $M$ or of spacetime points exists regardless. Hence there is no equivocation involved in the hole argument on whether the identity map plays a role.

Weatherall is correct that, \textit{as far as mathematical practice is concerned}, whenever we consider $\langle M,g\rangle$ and $\langle M, g'\rangle$ as isomorphic models, we do not assume any prior identity map on $M$. If we assume a prior identity map on $M$, then we are either not considering the two models isomorphic, or we are using the mathematical language improperly. The moral of his response is restricted to the mathematical part of the hole argument, namely that the mathematics does not force us to confront the metaphysical choice. If we want to infer a metaphysical dilemma from the formalism of general relativity, we must assume some metaphysical thesis to begin with.\footnote{ Note that Weatherall is neutral between substantivalism and relationalism (in general, metaphysical views of spacetime). Also, he does not reject (nor affirm) haecceities of \textit{physical} entities and that the same mathematical model can `represent two distinct physical situations' ([2018], p.4). }  But as the objection to Weatherall's approach goes, instead of solving the hole argument, it seems to only draw out the irrelevance of mathematical formalism to the argument.  (This irrelevance is expounded by Teitel [2019], who reconstructs the hole argument in terms of modality so that we can `directly discuss the phenomena we use this formalism to represent' (p.29).)

I very much agree with Weatherall's diagnosis of the misuse of mathematical formalism  in the hole argument (as it is standardly formulated) from the point of view of contemporary mathematical practice, which already incorporates the spirit of mathematical structuralism. I also agree with him that `our interpretations of our physical theories should be guided by the formalism of those theories' ([2018], p.2). I object to, however, the dichotomy between mathematics and physics (or metaphysics) implicitly assumed by authors on both sides. Instead of focusing on mathematical formalism in its narrow sense, we should focus on the entire logical framework that we use for expressing our worldviews, including defining concepts, formulating principles, constructing models, making deductions and so on. 
Accordingly, instead of using mathematical structuralism as a response to the hole argument, we should generalize the spirit to this whole framework in which scientific or fundamental discourse about our reality is best formulated. 
In particular, it ought not be the case that our best model in this framework is indeterminate between distinct physical situations, the difference of which is independently captured by a discourse external to this framework. In the case of Alice and Barbara, the apparent absurdity arises not because of a conflation of structural mathematical representations with physical situations, but because we are still using a meta-language that distinguishes between the two situations that are indistinguishable by structural features. 


This foundational approach, however, is not available to Weatherall because our standard formal framework lacks the technical means to express how isomorphic models ought to be interpreted and whether they refer to the same physical situation. The claim that  isomorphic models have the same representation capability is a meta-mathematical or meta-formal characterization of our mathematical practice. In particular, it is neither an axiom nor derivable from the axioms of our standard foundation. (In fact, it is not only unprovable in our standard set-theoretic foundation but also in category theory, which is worth mentioning because many consider the latter as the backdrop of contemporary mathematical practice. Although it is often said that category theory does not `care' about how many isomorphic copies of an object there are, it is nevertheless entirely possible to define multiple isomorphic copies.  As a simple example, we can define a category with two objects, each of which has an arrow pointing towards the other. The two objects, though structurally identical, are not the same object. This particular category is also not the same as the category of only one object that only has an arrow pointing towards itself, even though their only difference is how many isomorphic copies of an object they contain. Thus, it is not internal to the framework that isomorphic models have the same representational capacity.) 

This is where UF comes to the rescue. As a foundation that implements ontic structuralism (or so I have argued), UF generalizes the spirit of mathematical structuralism to the description of our physical world. Since we can develop everything we need in the foundational framework, there does not need to be a split between a mathematical structuralist language and a scientific one.  Moreover, insofar as modal claims (and more generally, metaphysical claims) are driven by science and used to express our worldview, they should also be rooted in our best foundational framework. For example, upon adopting UF, if we consider certain possibilities as structurally isomorphic, then \textsc{Univalence} is applicable and would entail that these possibilities are one and the same. Thus, we cannot carve out the space of metaphysical possibilities differently from how we view the space of best mathematical representations.

To put it in another way,  the UF-implemented ontic structuralism is both a metaphysical solution to the hole argument in denying the primitive identity of spacetime points, and also a mathematical solution in denying that there are distinct isomorphic models.\footnote{Note that `the UF-implemented ontic structuralism' amounts to the realist interpretation of UF as an implementation of ontic structuralism understood as \textsc{the Identity Thesis}, not just UF as a formal framework.} This, I think, is a  unification between formal treatment and ontic considerations that we need for solving the hole argument satisfactorily. Without the latter, a formal apparatus risks being irrelevant to the problem at hand, and without the former, a metaphysical position lacks clarity and risks being irrelevant to systematic sciences.

\paragraph{Generalization to (gauge) symmetries}

Diffeomorphism invariance concerned by the hole argument is usually subsumed under \textit{gauge invariance}. It is standardly considered as a formal symmetry, namely it relates models representing the same physical situation, as opposed to empirical symmetries that relate distinct but indistinguishable situations. The univalence solution to the hole argument can be applied to gauge symmetries in general, identifying all symmetry-related models. A well-known obstacle for this generalization is that symmetries are often \textit{not} isomorphisms.\footnote{Consider the simple example of a universal velocity boost in Newtonian mechanics. Let  $\langle E, \mathcal{\phi}\rangle$, $\langle E, \mathcal{\psi}\rangle$ be models related by this velocity boost, where $E$ is a Euclidean space, and $\phi,\psi$, with $\psi=\phi + b$ $(b\in \mathbb{R}\neq 0)$, are functions from $E$ to vectors representing the velocities of point particles. The two models are not isomorphic since except for very specific configurations of the velocities, there is no  automorphism $f$ on $E$ that preserves the velocities (that is, $\not\exists f(\forall x\in E)\psi(f(x))=f'(x)(\phi(x))$, where $f'$ is the derivative of $f$).} Nevertheless, I have argued elsewhere that we can reformulate any symmetry-related models as isomorphic (see also Dewar [2019b], Weatherall [2017], and Wallace [2019]).\footnote{\label{natural}Let me very briefly explain this trick, which involves using the tool of natural operators from category theory. This is a technique for finding out algebraic structures given all structure-respecting transformations. Very roughly, we first construct a category of models with symmetry transformations as morphisms. Then, we can derive the algebraic structures as natural transformations between functors from this category to the category of sets (see also Kolar et al [1993], Chen and Fritz [2021]). Note that Dewar ([2019b]) suggests that we can declare \textit{by fiat} that the given symmetries are isomorphisms. But this is not mathematically rigorous. The resulting objects are not necessarily algebraic structures, and we do not know what algebraic operators define them. Natural operators can remedy this problem.}  As a result, we can straightforwardly apply the univalence approach to gauge symmetries and conclude that all symmetry-related models are identical. Thus, in UF, the view about symmetries is completely clear-cut: all (perfect) symmetries are formal rather than empirical, since they represent the very same physical situation as enforced by \textsc{Univalence}.\footnote{Empirical symmetries can be understood as the restricted or imperfect kind relative to specific measurements that are unable to distinguish between situations represented by distinct non-isomorphic models (and relatedly, symmetries we observe before we formulate the relevant theories). For example, a very symmetric snowflake has rotational symmetries by $60^\circ$ with respect to our naked eye, but not with respect to a high-precision microscope.} It is intriguing to contemplate the similar spirit of UF and gauge theory (which introduces fields with gauge degrees of freedom): gauge theory takes advantage of the gauge degrees of freedom as representational `redundancy' while UF legitimizes this freedom (see, for example, Healey [2007] and Nguyen et al [2020] for related discussions of gauge theory).

 \section{Conclusion}
 
 I have shown that UF, featuring \textsc{Univalence}, is a rigorous foundational implementation of ontic structuralism with unique advantages over alternative proposals, and can shed light on important issues in philosophy of physics including the hole argument and promisingly the nature of symmetries in general.

\newpage
\appendix

\section{Set theory in UF}

Thinking in terms of sets is deeply entrenched in philosophers in the analytic tradition. It is therefore helpful to see how set theory can be modeled in HoTT/UF, and how it is different from standard set theory. But this is a large topic, so I will only sketch some very basic formal features here that help illustrate their structuralist characteristics. The presentation is based on UFP ([2013])

	Let us start with the definition of sets in HoTT.
	
	\begin{definition}[Set]
	 A type $A$ is a set iff for all $x,y:A$, and for all $p,q:x=y$, we have $p=q$.
	  More formally:
			\[\textbf{isSet}(A) :\equiv\Pi_{x,y:A}\Pi_{p,q:x=y}p=q.\]
	\end{definition}
 	We also need the basic notion of inductive types:
			
		\begin{definition}[Inductive Type]
			An inductive type X is a type that is generated by a finite set of constructors, each of which is a function  with	codomain X. This includes functions of zero arguments, which are simply elements of X.
		
		\end{definition}
In other words, an inductive type is defined through positing constant elements of the type and functions that map between elements of the type.
	
	\begin{example}
		The empty type \textbf{0}, which is an inductive type with no constructor, is a set.
	\end{example}
	Since there is no constructor, the proof is trivial.
	
\begin{example}
	The unit type \textbf{1}, which has one constant $\star:$ \textbf{1}, is a set.
\end{example}
\emph{Proof (sketch).} We need to show that 	\textbf{isSet}(\textbf{1}) $:\equiv$ $\Pi_{x,y:\textbf{1}}\Pi_{p,q:x=y}p=q$. It may be tempting to first reduce the desired theorem to $\Pi_{p,q:\star=\star}p=q$, and then prove it by induction on $p$ and $q$. But this is not possible because the type $\star=\star$ is not inductive and we would be stuck. Instead, we directly show that $\Pi_{x,y:\textbf{1}}x=y$ has only one element.  To establish this, we need to construct a bijection between it and \textbf{1}. Since \textbf{1} has only one element, let the function $f$ from $\Pi_{x,y:\textbf{1}}x=y$ to \textbf{1} send everything to $\star$, and let the function $g$ from \textbf{1} send $\star$ to \textbf{Refl$_\star$} (note: \textbf{Refl} is the trivial self-identity that holds of every element). To check that the two functions are inverses of each other, we need to show that $f(g(\star))=\star$ and $g(f(p))=p$ for all $p:\Pi_{x,y:\textbf{1}}x=y$. The first is trivial. The second is also straightforward because $g(f(p))=g(\star)=\textbf{Refl}_\star$, and we have that $p=\textbf{Refl}_\star$ for all $p:\Pi_{x,y:\textbf{1}}x=y$ by path induction. So we have shown that $\Pi_{x,y:\textbf{1}}x=y$ has only one element, and hence \textbf{1} is a set.

Now that we see sets are types whose elements have at most one way of identification, we can ask how many ways of identification a set has with itself. The answer is to be expected: for an $n$-element set, it has $n!$ ways of identification with itself which is the number of permutations of the elements. 
\begin{example}
		The boolean type \textbf{2}, an inductive type with two constants $0_2,1_2:$ \textbf{2}, is a set and has two ways of identification with itself.
\end{example}
Proof omitted. Informally, the two ways of self-identification are the identity and the swap, which can be shown to be distinct. That is, they map each element to different ones in \textbf{2}.

Sets in HoTT/UF are different from those in standard set theory in that their elements do not have primitive identity. To appreciate the difference, let's consider how union is defined in HoTT/UF. For example, consider two $n$-element sets $A,B$ that are not judgmentally equal. They are identified in UF since they are equinumerous.   If the notion of union were understood exactly as in standard set theory,  the union of $A$ with itself would be distinct from the union of $A$ with $B$, which would violate path induction (which implements the indiscernibility of identicals). To avoid this problem, the set-theoretic union in HoTT/UF is only definable for sets that are considered as subsets of a larger set. When we take set-theoretic union of two sets, we must specify how these sets are included in a larger set.  For a formal definition, we first need the notion of \textit{coproduct}, which corresponds to the notion of disjoint union in set theory. 

\begin{definition}[coproduct]
	For any types $A$ and $B$, the coproduct $A+B$ is the inductive type with two generating elements: left inclusion $i_1:A\rightarrow A+B$ and right inclusion $i_2:B\rightarrow A+B$.
\end{definition}
The induction principle for $A+B$ says that for any type $C$, if we have $f:A\rightarrow C$ and $g:B\rightarrow C$, then there is a function $h:A+B\rightarrow C$ such that $h\circ i_1=f$ and $h\circ i_2=g$. For an illuminating example, consider $A+A$. In set theory, we first force the two copies of $A$ to be disjoint, and then take the union of them (e.g., we can define $A+B$ to be $A\times \{0\}\cup B\times\{1\}$). The difference is important because it illustrates the structuralist spirit of HoTT/UF where the individuality of set elements is not primitive but rather derived from the structure of the whole. To continue:

\begin{definition}[Subset]
	$\{x:A \mid P(x)\}:\equiv \Sigma_{x:A} P(x)$ is a subset of $A$ if $A$ is a set and $P(x)$ is a proposition for all $x:A$.
\end{definition}

\begin{definition}[Disjunction]
	For propositions $P,Q$, their disjunction is $P\vee Q:\equiv ||P+Q||$, where $||P+Q||$ is the propositional truncation of $P+Q$ (a propositional truncation of a type adds identification between all elements of the type). 
\end{definition}

\begin{definition}[Union]
	$\{x:A \mid P(x)\}\cup \{x:A \mid Q(x)\}:\equiv \{x:A \mid P(x)\vee Q(x)\} $
\end{definition}

It is easy to check that the union of a subset with itself is just itself, and that two sets that are equinumerous are not necessarily identical as subsets of a larger set. Thus we avoid the violation of indiscernibility of identicals.

I hope this very brief introduction to set theory in HoTT/UF has given a flavor of how we can model set theory without assuming the primitive individuality of set elements as well as shown some important differences between the two. 

\section{Structure identity principle}

Given its centrality for implementing \textsc{Identity}, it would be beneficial to look at how exactly the structure identity principle is formulated as a theorem of UF, even if we do not go into its proof. We will make use of notions from category theory, which is in turn formulated in HoTT/UF.  

We start with the notion of precategory in HoTT/UF, which corresponds to the notion of category in the standard approach. We will see soon why the term `precategory' instead of `category' is used. For brevity, I do not include the full definitions, which are lengthy and can be easily found in UFP ([2013]).

	\begin{definition}[Precategory]
		A \emph{precategory} A consists of the following:
		\begin{itemize}
			\item A type \( \text{A}_0 \) whose elements are called \emph{objects}. We write $a:A$ for $a:A_0$.
			\item For each pair of objects \( a, b : \text{A} \), a type \( \text{Hom}_A(x, y) \) whose elements are called morphisms from \( a \) to \( b \).
		\end{itemize}
		They satisfy the usual conditions, such as the composition rule for morphisms and the existence of identity morphisms. We write the identity morphism for $a:A$ as   $1_a:Hom_A(a,a)$.
	\end{definition}
The notion of isomorphism between objects is defined similarly as in standard category theory:
\begin{definition}[Isomorphism]
A	morphism f : hom$_A$(a, b) is an isomorphism if there is a morphism g :
	hom$_A$(b, a) such that g ◦ f = 1$_a$ and f ◦ g = 1$_b$. We write $f\simeq g$.
\end{definition}

Now,   a category is a precategory that satisfies the univalence axiom:
\begin{definition}[Category]
 A precategory $A$ is a category if for any $a,b:A$, $a\simeq b$ implies $a=b$. 
\end{definition}
It is attractive to formally identify isomorphic objects in a category, since it is common to talk about objects up to isomorphism in category theory. Thus we replace the old notion of category (now `precategory') with this notion. 

Let's turn to structures. Let $X$ be a precategory. 

\begin{definition}[(P,H)-structure]
	A notion of structure (P, H) over X consists of the following.
	
	\begin{itemize} 
		\item A type family P : X$_0$ → U. For each x : X$_0$ the elements of Px are called (P, H)-structures on x.
		
		\item For x, y : X$_0$, f : Hom$_X$(x, y) and $\alpha$ : Px, $\beta$ : Py, a mere proposition
		H$_{\alpha\beta}$( f ).
		If H$_{\alpha\beta}$( f ) is true, we say that f is a (P, H)-homomorphism from $\alpha$ to $\beta$.
	\end{itemize}
$H$ satisfies further conditions that amount to the usual properties of homomorphisms.
\end{definition}

\begin{definition}
A precategory of (P, H)-structures, A = Str$_{(P,H)}$(X), consists of the following:
\begin{itemize}
	\item  A$_0:\equiv \Sigma (x:X_0) Px$.
	
	\item For $(x, \alpha), (y, \beta) : A_0$, we define	$Hom_A((x, \alpha), (y, \beta)) :\equiv	\{f : x \to y\mid	H_{\alpha\beta}( f )\}$
\end{itemize}

\end{definition}
For example, $X$ can be the precategory of sets, and A can be the precategory of groups, or rings, or topological spaces that are built on sets.

Finally let's turn to the structure identity principle:
\begin{theorem}[The structure identity principle]
	If $X$ is a category, then $Str_{(P,H)}(X)$ is a category.
\end{theorem}
As a special case, $X$ can be the category of sets, and  $Str_{(P,H)}(X)$ can be the category of manifolds defined set-theoretically. This gives us what we need for identifying set-theoretic structures in UF.

\end{document}